\documentclass{aa}

\usepackage{graphicx}
\usepackage{txfonts}
\usepackage[colorlinks=true, citecolor=blue, linkcolor=blue]{hyperref}
\usepackage{natbib}
\begin{document} 

\bibpunct{(}{)}{;}{a}{}{,} 

   \title{AMICO galaxy clusters in KiDS-DR3: Cosmological constraints from angular power spectrum and correlation function}
   \titlerunning{Cosmological constraints from angular power spectrum and correlation function}
    
   \author{M. Romanello
          \inst{1,2}
          \and
          F. Marulli\inst{1,2,3}
          \and
          L. Moscardini\inst{1,2,3}
          \and
          G. F. Lesci\inst{1,2}
          \and
          B. Sartoris\inst{4,5}
          \and
          S. Contarini\inst{1,2,3}
          \and
          C. Giocoli\inst{1,2,3}
          \and
          S. Bardelli\inst{2}
          \and
          V. Busillo\inst{6,7,8}
          \and
          G. Castignani\inst{1,2} 
          \and
          G. Covone\inst{6,7,8}
          \and
          L. Ingoglia\inst{1}
          \and
          M. Maturi\inst{9,10}
          \and
          E. Puddu\inst{7}
          \and
          M. Radovich\inst{11}
          \and
          M. Roncarelli\inst{2}
          \and
          M. Sereno\inst{2,3}
          }

   \institute{Dipartimento di Fisica e Astronomia “A. Righi” - Alma Mater Studiorum Università di Bologna, via Piero Gobetti 93/2, 40129 Bologna, Italy
   \and
   INAF - Osservatorio di Astrofisica e Scienza dello Spazio di Bologna, via Piero Gobetti 93/3, 40129 Bologna, Italy 
   \and 
   INFN - Sezione di Bologna, Viale Berti Pichat 6/2, 40127 Bologna, Italy
   \and 
   Universitäts-Sternwarte München, Fakultät für Physik, Ludwig-Maximilians-Universität München, Scheinerstrasse 1, 81679 München, Germany
   \and 
   INAF - Osservatorio Astronomico di Trieste, Via G. B. Tiepolo 11, 34143 Trieste, Italy
   \and 
   Dipartimento di Fisica “E. Pancini”, Università di Napoli Federico II, C.U. di Monte Sant’Angelo, via Cintia, 80126 Napoli, Italy
   \and 
   INAF - Osservatorio Astronomico di Capodimonte, Salita Moiariello 16, 80131 Napoli, Italy
   \and
   INFN - Sezione di Napoli, via Cintia, 80126 Napoli, Italy
   \and
   Zentrum für Astronomie, Universität Heidelberg, Philosophenweg 12, 69120 Heidelberg, Germany
   \and 
   ITP, Universität Heidelberg, Philosophenweg 16, 69120 Heidelberg, Germany
   \and 
   INAF - Osservatorio Astronomico di Padova, vicolo dell’Osservatorio 5, 35122 Padova, Italy
\\\\
        \email{massimilia.romanell2@unibo.it}
             }


 
  \abstract
  {We study the tomographic clustering properties of the photometric cluster catalogue derived from the Third Data Release of the Kilo Degree Survey, focusing on the angular correlation function and its spherical harmonic counterpart, the angular power spectrum. We measure the angular correlation function and power spectrum from a sample of 5162 clusters, with an intrinsic richness $\lambda^*\geq 15$, in the photometric redshift range $z\in [0.1, 0.6]$, comparing our measurements with theoretical models, in the framework of the $\Lambda$-Cold Dark Matter cosmology. We perform a Monte Carlo Markov Chain analysis to constrain the cosmological parameters $\Omega_{\mathrm{m}}$, $\sigma_8$ and the structure growth parameter $S_8\equiv\sigma_8 \sqrt{\Omega_{\mathrm{m}}/0.3}$. We adopt Gaussian priors on the parameters of the mass-richness relation, based on the posterior distributions derived from a previous joint analysis of cluster counts and weak lensing mass measurements carried out with the same catalogue. From the angular correlation function, we obtain $\Omega_{\mathrm{m}}=0.32^{+0.05}_{-0.04}$, $\sigma_8=0.77^{+0.13}_{-0.09}$ and $S_8=0.80^{+0.08}_{-0.06}$, in agreement, within $1\sigma$, with 3D clustering result based on the same cluster sample and with existing complementary studies on other datasets. For the angular power spectrum, we check the validity of the Poissonian shot noise approximation, considering also the mode-mode coupling induced by the mask. We derive statistically consistent results, in particular $\Omega_{\mathrm{m}}=0.24^{+0.05}_{-0.04}$ and $S_8=0.93^{+0.11}_{-0.12}$, while the constraint on $\sigma_8$ alone is weaker with respect to the one provided by the angular correlation function, $\sigma_8=1.01^{+0.25}_{-0.17}$. Our results show that the 2D clustering from photometric cluster surveys can provide competitive cosmological constraints with respect to the full 3D clustering statistics, and can be successfully applied to ongoing and forthcoming spectro/photometric surveys.}


   \maketitle
%

\section{Introduction} The spatial properties of the large-scale structure (LSS) of the Universe have been recognized as key cosmological probes. According to the $\Lambda-$cold dark matter ($\Lambda$CDM) model, galaxy clusters are the largest gravitationally bound systems that emerge from the cosmic web of LSS \citep[e.g.][]{Kaiser84}. They trace peaks in the large-scale matter density field, produced by gravitational infall and hierarchical merging of dark matter haloes \citep{Bardeen86, Tormen98, Despali16}. Since their growth is related to the expansion rate of the Universe and to the underlying distribution of matter, cluster statistics represent a powerful tool to understand the structure formation process, to constrain the neutrino mass \citep[e.g.][]{Marulli11, Villaescusa-Navarro14, Roncarelli15}, to investigate the nature of dark matter and dark energy \citep[e.g.][]{Mantz08, Vikhlinin09, Marulli12, Sartoris16, Costanzi19, Moresco21, Lesci_counts, Lesci22} and of gravity itself \citep[e.g.][]{Marulli21}.\\
\indent Despite the fact that cluster catalogues contain usually a lower number of objects with respect to galaxy catalogues, cosmology with clusters presents a series of key advantages. Galaxy clusters are hosted by the most massive virialised haloes, so they are highly biased tracers, i.e. more clustered than galaxies \citep[e.g.][]{Mo96, Moscardini01, Sheth01, Hutsi10, Allen11, Moresco21}. Furthermore, thanks to their lower peculiar velocities, galaxy clusters are relatively less affected by nonlinear dynamics at small scales, in particular by the effect of incoherent motions within virialised structures, that generates the so-called fingers-of-God effect \citep[e.g.][]{Veropalumbo14, Sereno15, Marulli17}. The impact of redshift-space distorsion (RSD) is thus reduced, allowing us to simplify theoretical assumptions in the modelling of their clustering signal. \\ \indent During the past years, cluster catalogues have been constructed from observations at several wavelengths \citep{Allen11}, for example by exploiting the X-ray emission from the diffuse intracluster medium \citep[ICM;][]{Rosati02, Bohringer04, Pacaud16}, the millimeter Sunyaev-Zel’dovich effect produced by the inverse Compton scattering between the hot ICM electrons and the cosmic microwave background (CMB) photons \citep{Vanderlinde10, Planck11}, and the optical and near-infrared (IR) starlight emission from galaxies \citep{Eisenhardt08, Bellagamba18}. The importance of this multiwavelength approach relies on the possibility to relate different observables, accessible with spectro-photometric observations, to the total mass of clusters, mostly composed of dark matter. The existence of a so-called mass-observable scaling relation \citep{Okabe10, Allen11, Giodini13, Sereno15X, Bellagamba19, Sereno20, Giocoli21} represents a useful link between the
theoretical mass function and the distribution of clusters in the space of survey observables, and gives us the opportunity to predict the effective bias of the cluster sample \citep[e.g.][]{Branchini17, Lesci22}. \\
\indent In the last decades, the cosmic distribution of the LSS has been investigated in a progressively more accurate and precise way. Typically, measurements of clustering are based on some cosmological assumption for the redshift-distance relation, and require an appropriate reconstruction of the position of cosmic structures, which can be provided by spectroscopic redshift surveys, like the Sloan Digital Sky Survey \citep[SDSS; see][]{Tegmark04} or, more recently, the Baryon Oscillation Spectroscopic Survey \citep[BOSS; see][]{Tojeiro12} and the Dark Energy Spectroscopic Instrument Legacy Survey \citep[DESI; see][]{Qianjun2021}. However, spectroscopic surveys are time-consuming thus, in a given amount of observational time, they have a series of limitations in terms of sky coverage and numbers of detected objects. On the other hand, ongoing and future photometric surveys, like the Kilo Degree Survey \citep[KiDS; see][]{deJong17, Kuijken19}, the Dark Energy Survey \citep[DES; see][]{DES16}, the Hyper Suprime-Cam (HSC) Subaru Strategic Program \citep[HSC-SSP; see][]{Aihara18}, the Vera C. Rubin Observatory Legacy Survey of Space and Time \citep[LSST; see][]{LSST12} and the \textit{Euclid} mission \citep{Laureijs11, Scaramella14, Amendola18, Scaramella22}, will allow us to cover a wider area, imaging also faint sources at high $z$.\\
\indent One of the most powerful tools of modern cosmology is the analysis of the two-point correlation function. The simplest and historically first used point-process statistics are the two-point angular correlation function, in configuration space, and its harmonic-space counterpart, the angular power spectrum \citep[][]{Hauser73, Peebles73}. In principle, these two statistics bring the same cosmological information, though in practice they have different sensitivities to different scales, due to the finite sizes of real catalogues, and thus the limited range of scales that can be probed. One of their fundamental advantages with respect to the full 3D study is that we can measure the clustering signal from the angular position alone, without any cosmological assumption in converting redshifts to distances \citep{Asorey12, Salazar14}.\\ 
\indent The aim of this work is to perform a cosmological analysis based on the catalogue of galaxy clusters identified by the Adaptive Matched Identifier of Clustered Objects \citep[AMICO; see][]{Bellagamba18} algorithm from the third data release of the Kilo Degree Survey (KIDS-DR3), presented in \citet{Maturi19}. Here, the availability of photometric redshift measurements allows us to divide the catalogue in shells and to perform a \textit{tomographic} study, which can provide independent constraints relative to the 3D reconstruction.\\
\indent The current analysis has been performed with the \textsc{CosmoBolognaLib} \citep{MarulliCBL}\footnote{\url{https://gitlab.com/federicomarulli/CosmoBolognaLib}, V6.1. The new likelihood functions to model the angular correlation function and power spectrum will be released in the upcoming version of the libraries. }, a set of {\em free software} C++ and Python libraries that we used to manage cluster catalogues, to measure their statistical quantities and to perform the Bayesian inferences.\\ \indent This work is part of a series of papers which aims at exploiting distant clusters in KiDS for both cosmological \citep{Bellagamba19, Giocoli21, Ingoglia22, Lesci_counts, Lesci22, Busillo23} and astrophysical studies \citep{Radovich20, Puddu21}. \\
\indent The paper is organised as follows. In Sect. \ref{Data} we present the AMICO KiDS cluster catalogue. In Sects. \ref{wtheta} and \ref{Sect_angular_power_spectrum} we describe the
methods used to measure and model the cluster angular correlation function and power spectrum, respectively. In Sect. \ref{Results} we discuss the results of the cosmological analysis. Finally, in Sect. \ref{Conclusions} we draw our conclusions.   
\begin{figure*}[htbp]
    \centering
    \includegraphics[width=\textwidth]{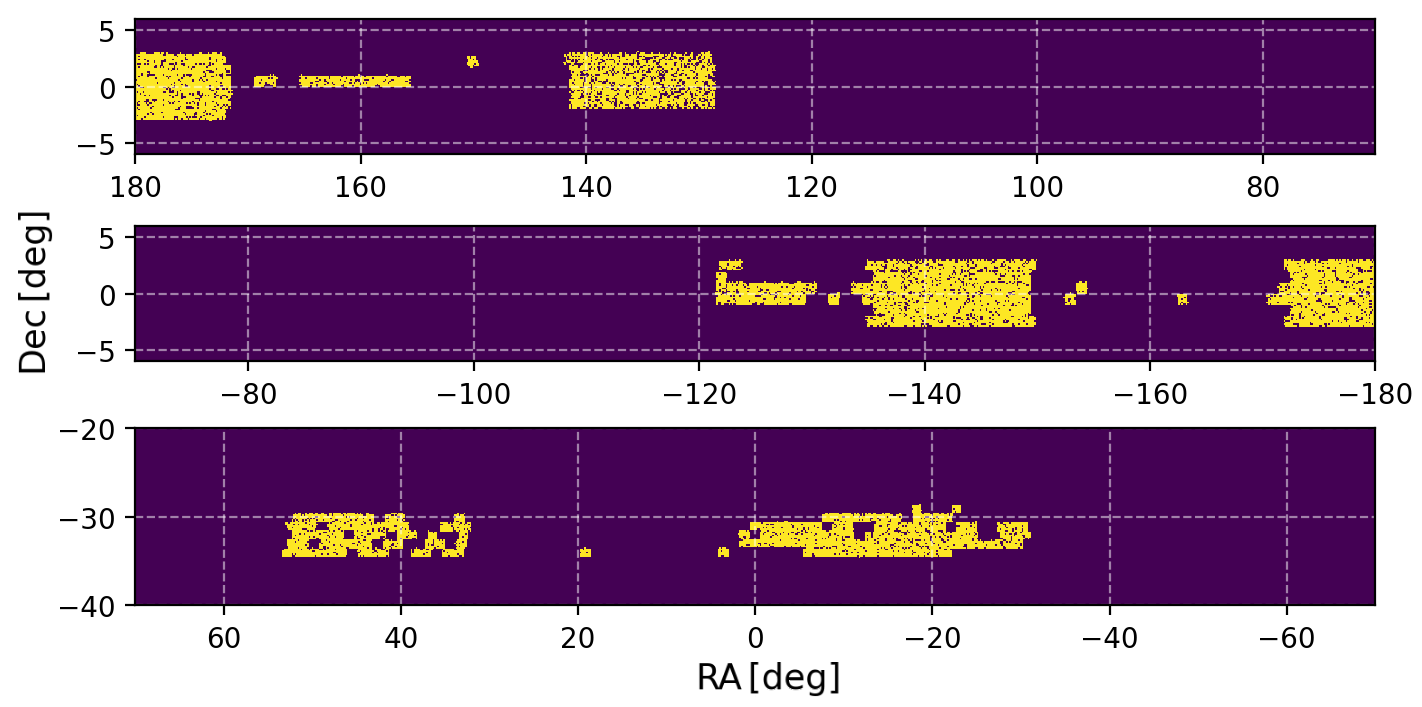}
      \caption{The KiDS-DR3 footprint binary mask \citep{Hildebrandt17}, pixelated with \textsc{Healpix} \citep{Gorski05}, with a resolution given by $N_{\mathrm{side}}=512$. The top and the middle panels cover the extension of KiDS-N, while the bottom panel refers to KiDS-S. Clusters are distributed within the unmasked area (yellow). Small holes and irregularities reflect the presence of severe incompleteness and substantial photometric degradation, due to satellite tracks, stars etc.}
    \label{Fig_mask}
   \end{figure*}
\section{Data: the AMICO KiDS-DR3 catalogue}
\label{Data}
KiDS is a European Southern Observatory (ESO) public optical imaging survey, obtained with the OmegaCAM wide-field imager \citep{Kuijken11} mounted on the 2.6m  Very Large Telescope (VLT) Survey Telescope (VST), at the Paranal Observatory. This work is focused on its third release, KiDS-DR3 \citep{deJong17}. The DR3 covers a total area of 438 deg$^2$ over two fields, one equatorial (KiDS-N) and the other near the South Galactic Pole (KiDS-S), with aperture photometry in \textit{u, g, r}, and \textit{i} bands down to the limiting magnitudes of 24.3, 25.1, 24.9, and 23.8, respectively. A final effective area of 377 deg$^2$ \citep{Maturi19}, displayed by the pixelated KiDS-DR3 footprint binary mask presented in Fig. \ref{Fig_mask} \citep{Hildebrandt17}, which can be found on the KiDS website\footnote{\url{https://kids.strw.leidenuniv.nl/DR3/lensing.php}}, was obtained after rejecting all regions affected by satellite tracks, within haloes produced by bright stars and within secondary/tertiary halo masks used for the weak lensing analysis \citep{deJong15, Kuijken15}.\\ \indent As discussed in \citet{Maturi19}, clusters were detected thanks to the application of the AMICO algorithm \citep{Bellagamba18}, which identifies galaxy overdensities by exploiting a linear matched optimal filter. Cluster detection relies only on angular coordinates, magnitudes, and photometric redshifts of galaxies, avoiding a colour-based selection related to the red-sequence of clusters.\\
\indent The complete sample contains 7988 objects, with a signal-to-noise ratio $(S/N) > 3.5$, in the  redshift range $z\in[0.10,0.80]$. We limit the current study to $z\in[0.10, 0.60]$ because our model is based on the mass-richness scaling relation estimated from the stacked weak-lensing analysis presented in \citet{Bellagamba19} and \citet{Lesci_counts}, which has been calibrated in this photo-$z$ range. The cluster detection algorithm returns an unbiased redshift estimate with respect to the input photometric catalogue, but it is sensitive to the photo-$z$ bias of the underlying galaxy population, discussed in \citet{deJong17} for the KiDS survey. Thus, as suggested by \citet{Maturi19}, we correct the estimated cluster redshifts with the relation $z=z_{\mathrm{est}}-0.02(1+z_{\mathrm{est}})$.\\ 
\indent The mass proxy for the scaling relation used in this paper is the intrinsic richness provided directly by AMICO, defined as the sum of the membership probabilities:
\begin{equation}
\label{intrinsic_richness}
    \lambda^*_ j=\sum_{i=1}^{N_{gal}} P_i(j) \quad \textrm{with} \quad \begin{cases}
m_i<m^*(z_j)+1.5 \\
r_i(j)<R_{200}(z_j),\\
\end{cases}
\end{equation}
where $P_i(j)$ is the probability assigned by AMICO to the $i$-th
galaxy of being a member of a given detection $j$, $r_i(j)$ is the distance of the $i$-th
galaxy from the centre of the $j$-th cluster, $R_{200}(z_j)$ is the sphere radius in which the mean density is 200 times the critical density at redshift $z_j$, and $m^*$ is a function of redshift representing the typical magnitude used in the Schechter function of the cluster model employed by the AMICO algorithm \citep{Maturi19}. Thus for a given detection the intrinsic richness represents the expected number of visible galaxies, under the condition expressed in Eq. \eqref{intrinsic_richness}. According to this definition, it is a nearly redshift-independent quantity because the threshold $m^* + 1.5$ is well below the magnitude limit of the galactic sample, in the considered redshift interval. For our analysis we select clusters with $\lambda^* \geq 15$, which ensures a purity higher than $97\%$ and a completeness higher than $50\%$ over the whole sample \citep{Maturi19}. Finally, to perform a tomographic analysis, we split our catalogue in three different redshift bins. Thinner shells preserve clustering information along the line of sight and are closer to a full 3D study \citep{Asorey12, Salazar14, Balaguera18}, but they reduce the projected number density of clusters, and thus the accuracy and the precision of the measurements \citep{Salazar14}. The redshift bin width is chosen to be five times larger than the maximum photometric error, while we increase the amplitude of the first redshift shell to improve the statistics. The final sample contains 5162 clusters, 1019 in $z \in (0.10, 0.30]$, 2072 in $z \in (0.30, 0.45]$ and 2071 in $z \in (0.45, 0.60]$.
\section{The angular correlation function of AMICO KiDS-DR3 catalogue}
In this section we describe the methods used to measure and model the angular correlation function.
   \begin{figure}
   \centering
   \includegraphics[width=\hsize]{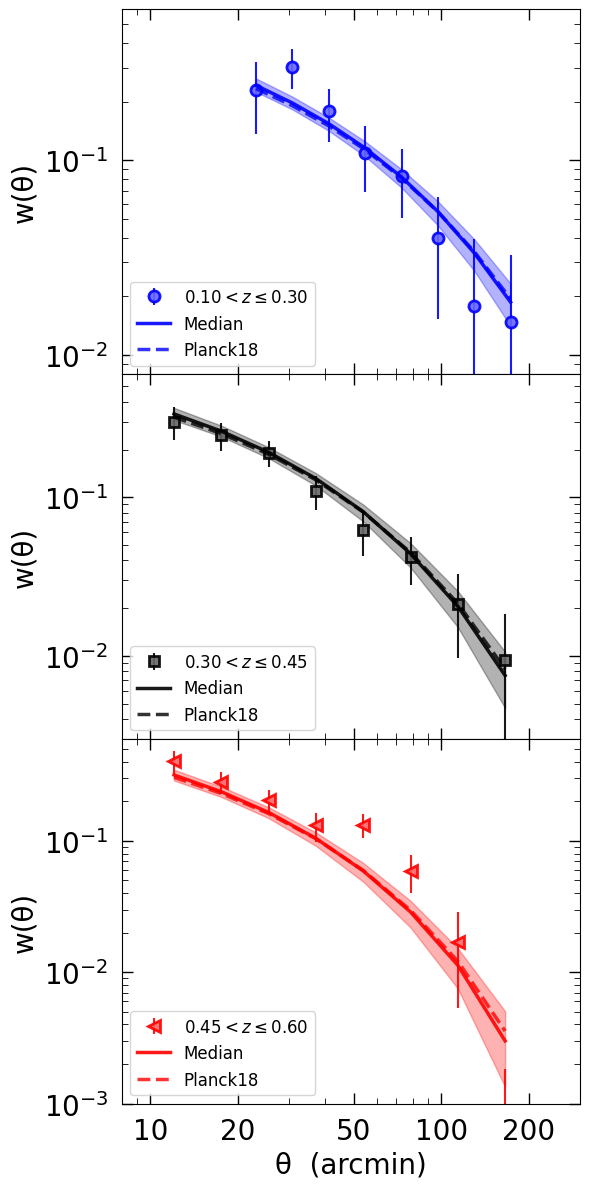}
      \caption{The angular correlation function measured in three redshift bins: $z\in(0.10, 0.30]$ (blue circles), $z\in(0.30, 0.45]$ (black squares), $z\in(0.45, 0.60]$ (red triangles). Error bars are estimated as the diagonal terms of the jackknife covariance matrix. The dashed lines represent the model computed with cosmological parameters by \citet[][Table 2, TT, TE and EE+lowE]{Planck20}. The solid lines show the median of the model distribution computed from the combined posterior of our cosmological analysis, while the shaded regions highlight the $68\%$ confidence levels.}
         \label{fig_wtheta_measure}
   \end{figure}
%

\label{wtheta}
\subsection{The angular correlation function estimator}
The joint probability of finding two clusters in the solid angle elements $\delta \Omega_1$ and $\delta \Omega_2$, at a distance $\theta$ is given
by: \begin{equation}
    \delta P(\theta)= n_{\Omega}^2[1+w(\theta)]\delta \Omega_1 \delta \Omega_2,
\end{equation}
where $n_\Omega$ is the mean number of clusters per unit solid angle. Thus, the angular correlation function $w(\theta)$ represents the excess probability of finding a pair of objects separated by the angular distance $\theta$, relative to that expected from a random distribution. We measured the observed angular correlation function using the \citet{Landy93} (LS) estimator:
\begin{equation}
    w_{\mathrm{LS}}(\theta)=\frac{DD(\theta)+RR(\theta)-2DR(\theta)}{RR(\theta)},
\end{equation}
where $DD(\theta)$, $RR(\theta)$, and $DR(\theta)$ are the number of data-data, random-random, and data-random pairs in the angular bin $\theta \pm \Delta \theta/2$, respectively. The measurement is performed in eight logarithmically-spaced bins, between 10-20 and 200 arcmin, with a conservative angular separation which takes into account, for the lower limit, the maximum virial cluster size in every redshift bin and, for the upper limit, the angular scale of the survey. The results are shown for each redshift bin in Fig. \ref{fig_wtheta_measure}, where they are also compared to the model presented in Sect. \ref{Sect_wtheta_model} and to the result of the cosmological analysis of Sect. \ref{Results}.  \\
\indent We construct the random catalogue by randomly extracting the angular (RA, Dec) cluster coordinates within the survey observational tiles, using the same masks adopted in \citet{Maturi19}. To limit shot noise effects, our random catalogue is 30 times larger than the original one. The covariance matrix is estimated through the jackknife method \citep{Norberg09}. Specifically, for each redshift slice, we project our catalogue onto the celestial sphere, using the equal-area \textsc{Healpix} pixelisation scheme \citep[see Sect. \ref{Sect_angular_power_spectrum_map}]{Gorski05}, with a low-resolution $N_{\mathrm{side}} = 128$, i.e. corresponding to a pixel side of 27 arcmin. Clusters belonging to the same pixel are considered part of a unique region. Therefore, the exact number of regions depends on the quantity of clusters available in each redshift bin and is of the order of 1000. This allows us to estimate the covariance matrix with different $N_\mathrm{JK}$ measurements of the angular correlation function, obtained after removing one region at a time.
\subsection{The angular correlation function model}
\label{Sect_wtheta_model}
On linear scales, the cluster density field, $\delta_{\mathrm{cl}}(\mathbf{x})$, is related to the dark matter density field, $\delta_{\mathrm{DM}}(\mathbf{x})$, through a scale-independent bias, $b_{\mathrm{cl}}(z)$: 
\begin{equation}
  \delta_{\mathrm{cl}}(\mathbf{x})\equiv\frac{n_{\mathrm{cl}}(\mathbf{x})-\Bar{n}_{\mathrm{cl}}}{\Bar{n}_{\mathrm{cl}}}= b_{\mathrm{cl}}(z)\delta_{\mathrm{DM}}(\mathbf{x}),  
\end{equation}
which implies that in Fourier space, $P(k)=b^2_{\mathrm{cl}}P_{\mathrm{DM}}(k)$, where $n_{\mathrm{cl}}(\mathbf{x})$ is the cluster density, $\Bar{n}_{\mathrm{cl}}$ is its mean value and $P_{\mathrm{DM}}(k)$ is the linear dark matter power spectrum. We employed the fitting formulae provided by \citet{Eisenstein98}, which in the angular range of our $w(\theta)$ and $C_\ell$ analyses produce consistent results to those provided by CAMB \citep{Lewis00} and CLASS \citep{CLASSI, CLASSII}. Given the normalised selection function $\phi^i(z)$ in the $i$-th photometric redshift bin (see Sect. \ref{Selection function}), we can project the density field onto the celestial sphere, in a given direction $\hat{\mathbf{n}}$ on the sky:
\begin{equation} 
\label{Eq_delta_cl}\delta_{\mathrm{cl}}^i(\hat{\mathbf{n}})=\int \mathrm{d}z \phi^i(z) \delta_{\mathrm{cl}}^i (\mathbf{x}).
\end{equation}
The angular correlation function at a given separation $\theta$ is the projection of the 3D spatial correlation function, $\xi(s)$:
\begin{equation}
\label{Eq_wtheta_theor}
    w^{ij}(\theta)=\int \int \mathrm{d}z_1 \mathrm{d}z_2 \phi^i(z_1) \phi^j(z_2) \xi(s),
\end{equation}
where $s=\sqrt{r^2(z_1)+r^2(z_2)-2r(z_1)r(z_2)\cos\theta}$ and $r(z)$ is the comoving distance at redshift $z$. 
In this work, we assume the plane-parallel approximation and we parameterise the linear power spectrum in redshift space as:
\begin{equation}
    P(k, \mu)=(b_{\mathrm{eff}}+f \mu^2)^2 P_{\mathrm{DM}}(k),
\end{equation}
where $f\equiv \frac{\mathrm{d}\ln D}{\mathrm{d} \ln a}$ is the linear growth rate, $\mu$ is the cosine of the angle between $\textbf{k}$ and the line of sight and $b_{\mathrm{eff}}$ is the effective bias, i.e. the halo bias weighted with the halo mass function (see Sect. \ref{Effective_bias}). The Fourier transform of the power spectrum gives us the 3D correlation function, which can be expressed in terms of multipoles $\xi_l(s)$ and Legendre polynomials $P_l(\mu)$ \citep{Hamilton92}:
\begin{equation}
    \xi(s, \mu)=\xi_0(s)+\xi_2(s)P_2(\mu)+\xi_4(s)P_4(\mu)+\mathcal{O}(s^4).
\end{equation}
We keep only the monopole because it includes most of the information \citep{Salazar14, GarciaFarieta20}. It can be written as a function of the real-space correlation function, $\xi_{\mathrm{DM}}(r)$:
\begin{equation}
    \xi_0(s)=\left[b^2_{\mathrm{eff}}+\frac{2}{3}b_{\mathrm{eff}}f+\frac{1}{5}f^2 \right]\xi_{\mathrm{DM}}(r).
\end{equation}
\subsection{Redshift selection function model}
\label{Selection function}
Photometric redshifts have larger uncertainties than spectroscopic ones.
Because of photo-$z$ errors, cluster photometric redshift distributions can be different from the true redshift distributions, thus we need to account for the conditional probability of having a cluster at the true redshift, $z$, given the observed redshift, $z_{\mathrm{phot}}$. 
\begin{figure}
   \centering
   \includegraphics[width=\hsize]{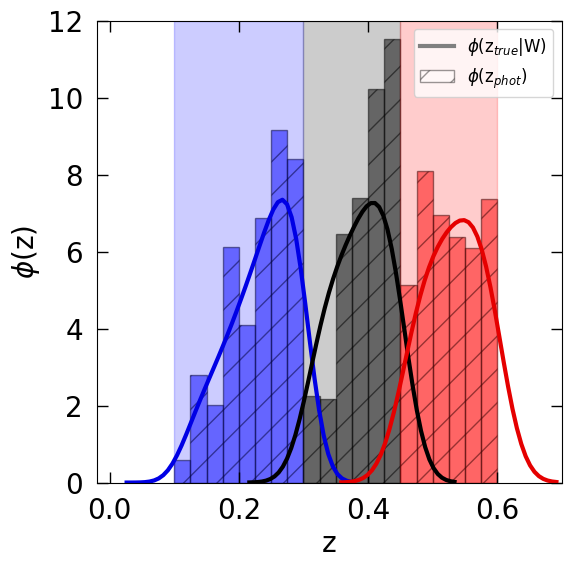}
      \caption{Normalised redshift distributions. The histograms represent $\phi(z_{\mathrm{phot}})$, from the photometric cluster catalogue, while $\phi(z_{\mathrm{true}}|W)$ is predicted from Eq. \eqref{eq_selection_function}, using the theoretical halo mass function model by \citet{Tinker08}, with cosmological parameters provided by \citet[][Table 2, TT, TE and EE+lowE]{Planck20}, and convolved with our photometric window function, $W$. The shaded areas indicate the limits of our photometric redshift bins.  
      }
         \label{Fig_redshift_distribution}
   \end{figure}
   
In Eqs. \eqref{Eq_delta_cl} and \eqref{Eq_wtheta_theor} we consider the radial selection function as the normalised cluster distribution in a given redshift bin $\Delta z_i$. In other words, it represents the probability of including a cluster in the corresponding photometric shell, depending on the selection characteristics of our study, namely on the binning strategy \citep{Asorey12}. 
Our photometric volume-limited survey is selected by the top-hat window function:
    \begin{equation}
    W(z_{\mathrm{phot}})=  \begin{cases}
    0 \qquad  z_{\mathrm{phot}}\leq z^i_{\mathrm{min}} \ \mathrm{or} \ z_{\mathrm{phot}}>z^i_{\mathrm{max}} \\
    1 \qquad  z^i_{\mathrm{min}}< z_{\mathrm{phot}} \leq z^i_{\mathrm{max}} \\
\end{cases}
,
\end{equation}
where $z_{\mathrm{min}}$ and $z_{\mathrm{max}}$ represent the lower and the upper limits of our photometric redshift interval, respectively. Including objects into redshift shells of a given redshift width $\Delta z_i$ allows us to `integrate out' the effect of photo-$z$ errors \citep{Bykov23}. The conditional probability of detecting a cluster in a sample selected with this window function, i.e. our normalised redshift distribution, is obtained with the following convolution \citep{Budavari03, Crocce11, Hutsi14}:
\begin{equation}
\label{eq_selection_function}
    \phi^i(z)=\phi(z|W)=\phi(z)\int_0^\infty \mathrm{d}z_{\mathrm{phot}}W(z_{\mathrm{phot}})P(z_{\mathrm{phot}}|z),
\end{equation}
where $\phi(z)$ is the true underlying full redshift distribution:
\begin{equation}
    \phi(z)=\frac{\frac{\mathrm{d}N}{\mathrm{d}z}}{\int \mathrm{d}z \frac{\mathrm{d}N}{\mathrm{d}z}}=\frac{1}{N}\frac{\mathrm{d}N}{\mathrm{d}z},
\end{equation}
and $P(z_{\mathrm{phot}}|z)$ is a Gaussian distribution, as derived by \citet{Lesci_counts, Lesci22} from the mock catalogues described in \citet{Maturi19}, whose mean is $z$, while the standard deviation is equal to:
\begin{equation}
    \sigma_z=\sigma_{z,0}(1+z),
\end{equation}
with $\sigma_{z,0} = 0.02$. An estimate of the true redshift distribution can be then obtained as follows:
\begin{equation}
    \frac{\mathrm{d}N}{\mathrm{d}z}=\Omega_{\mathrm{sky}} \frac{\mathrm{d}V}{\mathrm{d}z \mathrm{d} \Omega} \int_{0}^{\infty} \frac{\mathrm{d}n(M, z)}{\mathrm{d}M} \mathrm{d}M
    \int_0^\infty \mathrm{d}\lambda^*P(\lambda^*|M, z),
\end{equation}
 where $\Omega_{\mathrm{sky}}$ is the survey area in rad$^2$. Here, the cosmological dependence is provided by $\frac{\mathrm{d}V}{\mathrm{d}z \mathrm{d} \Omega}$, the derivative of the comoving volume, by $\frac{\mathrm{d}n(M,z)}{\mathrm{d}M}$, the halo mass function modelled with the functional form by \citet{Tinker08}, and by $P(\lambda^*| M, z)$, where $\lambda^*$ is the intrinsic richness. The latter integral convolves the theoretical mass function, taking into account the shape of the mass-observable scaling relation of the cluster sample \citep{Lesci_counts}:
 \begin{equation}
     P(\lambda^*|M, z)= \frac{P(M|\lambda^*, z)P(\lambda^*|z)}{P(M|z)}, 
 \end{equation}
 where $P(\lambda^*|z)$ is a cosmology-independent power-law with an exponential cut-off calibrated from mock catalogues \citep{Lesci_counts}, $P(M|z)$ is a normalisation factor computed as the integral of the numerator, $P(M|\lambda^*,z)$ is modelled as a log-normal distribution, in which the mean is given by the mass-observable scaling relation and the root mean square is a free parameter of the model: \begin{equation}
     P(\log M| \lambda^*,z) =\frac{1}{\sqrt{2\pi}\sigma_{\mathrm{intr}}}\mathrm{exp} \left(-\frac{x^2(M, \lambda^*, z)}{2 \sigma_{\mathrm{intr}}^2}\right).
 \end{equation}
Here, 
\begin{multline}
    x(M, \lambda^*, z)=\log \frac{M}{10^{14} M_\odot/h}-\\- \left(\alpha+\beta\log \frac{\lambda^*}{\lambda^*_{\mathrm{piv}}}+\gamma\log \frac{E(z)}{E(z_{\mathrm{piv}})} \right),
\end{multline}
where $E(z)\equiv H(z)/H_0$. We set $\lambda_{\mathrm{piv}}=30$ and $z_{\mathrm{piv}}=0.35$, which represent the central values of intrinsic richness and redshift in the ranges covered by the whole sample, as found by \citet{Bellagamba19}. The intrinsic scatter is modelled with two free parameters, $\sigma_{\mathrm{intr},0}$ and $\sigma_{\mathrm{intr}, \lambda^*}$, as follows:
\begin{equation}
    \sigma_{\mathrm{intr}}=\sigma_{\mathrm{intr},0}+\sigma_{\mathrm{intr}, \lambda^*}\log \left( \frac{\lambda^*}{\lambda^*_{\mathrm{piv}}}\right).
\end{equation}
Finally, we need to compute the cluster redshift distribution in a given redshift bin, accounting also for the probability of measuring $\lambda^*_{\mathrm{obs}}$ given the true intrinsic richness $\lambda^*$, and modelling the selection effects and the incompleteness of the sample. This requires a further convolution in the intrinsic richess bin $\Delta \lambda^*_i=\Delta \lambda^*(\Delta z_i)$, with a Gaussian $P(\lambda^*_{\mathrm{obs}}|\lambda^*)$. We considered an uncertainty of $17\%$ on $\lambda^*_{\mathrm{obs}}$, as in \citet{Lesci_counts, Lesci22}. In Fig. \ref{Fig_redshift_distribution} we show the KiDS-DR3 photometric redshift distribution and the complete number counts model, expressed by: 
\begin{equation}
\label{eq_dN_dzi}
\begin{split}
   \frac{dN}{dz^i}=\Omega_{\mathrm{sky}} \frac{\mathrm{d}V}{\mathrm{d}z \mathrm{d} \Omega} \int_{0}^{\infty} \mathrm{d}M \frac{\mathrm{d}n(M, z)}{\mathrm{d}M}
    \int_0^\infty \mathrm{d}\lambda^*P(\lambda^*|M, z) \\ \int_{\Delta z_i} \mathrm{d} z_{\mathrm{phot}}P(z_{\mathrm{phot}}|z)
    \int_{\Delta \lambda^*_i} \mathrm{d} \lambda^*_{\mathrm{obs}}P(\lambda^*_{\mathrm{obs}}|\lambda^*). 
\end{split}
\end{equation}
\subsection{Effective bias of the cluster sample}
\label{Effective_bias}
Under the assumption that clusters trace the locations of the large-scale dark matter haloes, we adopt a simple linear relation between dark matter and cluster power spectra: $b^2_{\mathrm{cl}}=P(k)/P_{\mathrm{DM}}(k)$. While in general the bias is expected to be a function of scale, at large linear scales it can be considered as a scale-independent quantity \citep[e.g.][]{Estrada09, Manera11, Sawangwit11}. On the other hand, the  evolution of both dark matter and cluster clustering leads to a bias which is a function of halo mass and redshift \citep{Estrada09}. \\ 
\indent Following the approach of \citet{Lesci22}, we assume a constant bias, neglecting any redshift evolution within the broad photometric shell. The effective bias is derived theoretically, as the average over the number of clusters $N_i$ in the photometric sample:
\begin{equation}
\begin{split}
    \label{eq_bias_eff}b^i_{\mathrm{eff}}=\frac{1}{N_i}\sum^{N_i}_{j=1}\int_0^\infty\mathrm{d}z \int_0^\infty \mathrm{d}\lambda^*\int_0^\infty \mathrm{d}M b(M,z) P(M|\lambda^*,z) \times \\ \times P(z|z_{\mathrm{phot},j})
    P(\lambda|\lambda_{\mathrm{obs},j}),
\end{split}
\end{equation}
where $b(M,z)$ is the halo-bias according to the model presented in \citet{Tinker10}. 
\begin{table*}[h!]
\caption{Parameters, prior and posterior mean and percentiles of the cosmological analyses.}        
\label{table_par}      
\centering          
\begin{tabular}{c c c c}     
\hline\hline       
Parameter & Description & Prior $w(\theta)$ - $C_\ell$ & Posterior $w(\theta)$ - $C_\ell$ \\ 
\hline                    
    $\Omega_{\mathrm{m}}$  & Total matter density parameter & $[0.1, 0.7]$ & $0.32^{+0.05}_{-0.04}$ - $0.24^{+0.05}_{-0.04}$\\
    
    $\sigma_8$  & Amplitude of the power spectrum on the scale of 8 $h^{-1}$ Mpc & $[0.3, 1.5]$ & $0.77^{+0.13}_{-0.09}$ - $1.01^{+0.25}_{-0.17}$\\
    
    $S_8\equiv \sigma_8(\Omega_{\mathrm{m}}/0.3)^{0.5}$& Structure growth parameter & - & $0.80^{+0.08}_{-0.06}$ - $0.93^{+0.11}_{-0.12}$\\
    
    $\Omega_\mathrm{b}$  & Baryon density parameter & $\mathcal{G}(0.0486, 0.0017)$ & -\\

    $n_\mathrm{s}$  & Primordial spectral index & $\mathcal{G}(0.9649, 0.0210)$ & -\\

    $h\equiv H_0/(100 \, \mathrm{km/s/Mpc})$  & Normalised Hubble constant & $\mathcal{G}(0.7, 0.1)$ & -\\
    
    $\alpha$  & Normalisation of the mass-richness scaling relation & $\mathcal{G}(0.04, 0.04)$ & -\\

    $\beta$  & Slope of the mass-richness scaling relation & $\mathcal{G}(1.72, 0.08)$ & -\\

    $\gamma$  & Redshift evolution of the mass-richness scaling relation & $\mathcal{G}(-2.37, 0.40)$ & -\\

    $\sigma_{\mathrm{intr},0}$  & Normalisation of $\sigma_{\mathrm{intr}}$ & $\mathcal{G}(0.18, 0.09)$ & -\\

    $\sigma_{\mathrm{intr},\lambda^*}$  & $\lambda^*$ evolution of $\sigma_{\mathrm{intr}}$ & $\mathcal{G}(0.11, 0.20)$ & -\\

    $\mathcal{S}_1$  & Extra shot noise bin $1$ & - $\;\mathcal{G}(0.0, 4.2 \times 10^{-6})$ & -\\

    $\mathcal{S}_2$  & Extra shot noise bin $2$ & -   $\;\mathcal{G}(0.0, 2.0 \times 10^{-6})$ & -\\

    $\mathcal{S}_3$  & Extra shot noise bin $3$ & -   $\;\mathcal{G}(0.0, 2.2 \times 10^{-6})$ & -\\
    
\hline                  
\end{tabular}
\end{table*}
\section{The angular power spectrum of AMICO KiDS-DR3 catalogue}
\label{Sect_angular_power_spectrum}
In this section we describe the methods used to measure and model the angular power spectrum.
\subsection{The pixelated density map}
\label{Sect_angular_power_spectrum_map}
For every different redshift bin, we generate a cluster density map by projecting the catalogue objects onto the celestial sphere, using the \textsc{Healpix} pixelisation \citep{Gorski05}, with a resolution $N_{\mathrm{side}} = 512$, which ensures a pixel size of approximately 7 arcmin, comparable with the minimum angular scale exploited in the $w(\theta)$ analysis. The density contrast $\delta_{\mathrm{cl}, I}$ in each pixel $I$ is given by:
\begin{equation}
    \delta_{\mathrm{cl}, I}=\frac{n_{\mathrm{cl},I}}{\Bar{n}_{\mathrm{cl}}}-1,
\end{equation} where $n_{\mathrm{cl},I}$ is the cluster number density in the $I$-th pixel and $\Bar{n}_{\mathrm{cl}}$ is its mean, computed in the unmasked area of the survey. The pixelisation procedure smooths information on scales smaller than the pixel size, i.e. $\ell>1500$, which is, in our case, well above the maximum value of $\ell$ used in our analysis. Different observational effects, like stellar density, air-mass, sky flux and reddening might introduce biases in the galaxy photometry, if not taken into account \citep{Loureiro19}. This could lead to biases in the cluster detection, thus in the clustering statistics. Anyway, regions affected by these effects are already excluded, since we estimate the cluster density in the unmasked field only, given by the KiDS-DR3 footprint binary mask. Furthermore, as \citet{Maturi19} imposed the strict magnitude cut at $r=24$ (note that the limit in the $r$ band for KiDS-DR3 is 24.9, see Sect. \ref{Data}), corresponding to the depth of the shallowest tile, the mean cluster count does not depend on the sky position \citep{Lesci22}. For these reasons, following for example \citet{Branchini17}, clusters are counted as single objects, thus no weighting scheme has been applied to account for their selection effects. In Fig. \ref{Fig_mask} we show the KiDS-DR3 footprint mask, degraded from a resolution of $N_{\mathrm{side}}=2048$ to $N_{\mathrm{side}}=512$, keeping the $377$ deg$^2$ survey area, which yields to a final unmasked sky fraction of $f_{\mathrm{sky}}\approx 0.9\%$. The irregular geometry of KiDS-DR3 survey is reflected in the large number of small holes contained in the mask. Thus, we avoid any apodisation of the mask, which would lead to a significant modification in the shape and to a non-trivial loss of area \citep{White22}. 
\subsection{The angular power spectrum estimator}
\begin{figure}
   \centering
   \includegraphics[width=\hsize]{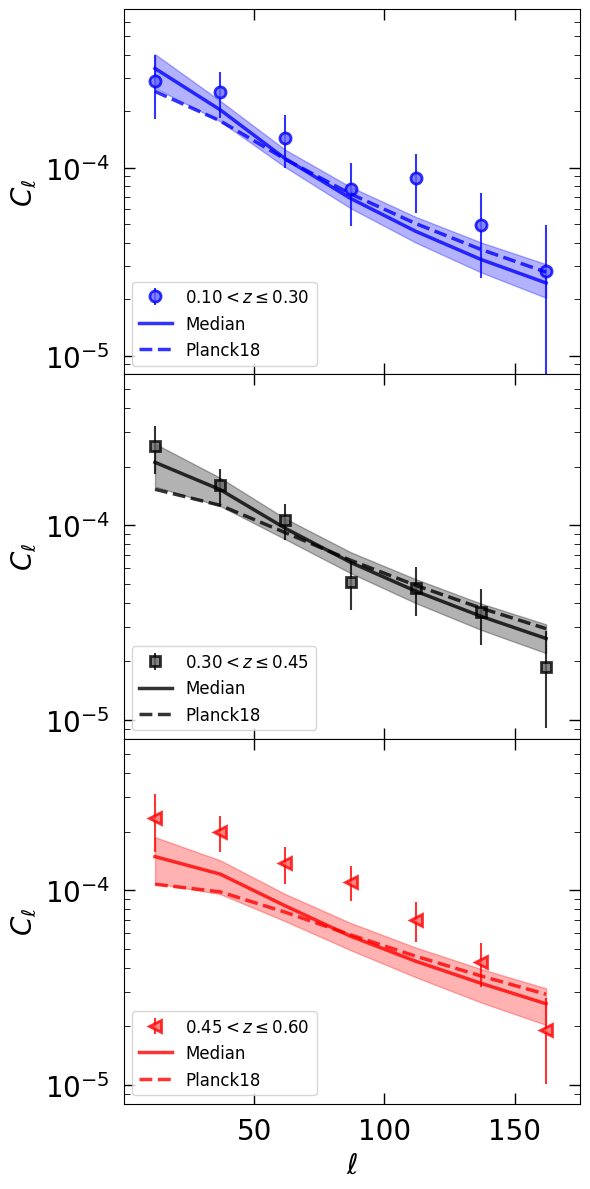}
      \caption{The angular power spectrum measured in three redshift bins: $z\in(0.10, 0.30]$ (blue circles), $z\in(0.30, 0.45]$ (black squares), $z\in(0.45, 0.60]$ (red triangles). Error bars are estimated as the diagonal terms of the jackknife covariance matrix. The dashed lines represent the model computed with cosmological parameters by \citet[][Table 2, TT, TE and EE+lowE]{Planck20}. The solid lines show the median of the model distribution computed from the combined posterior of our cosmological analysis, while the shaded regions highlight the $68\%$ confidence levels.}
      \label{fig_cl_measure}
   \end{figure}
\begin{figure}
   \centering
   \includegraphics[width=\hsize]{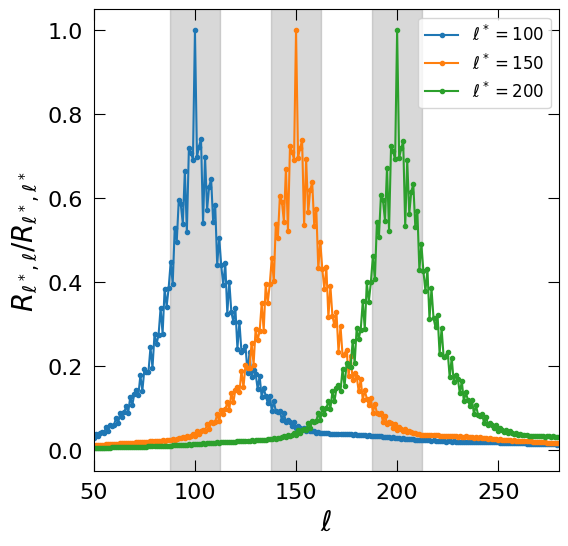}
      \caption{Normalised elements of the mixing matrix $R_{\ell \ell'}$, centred in three different multipoles $\ell^*=100, 150, 200$. The functions decrease as we move away from $\ell^*$. Their value give us a quantitative amount of the correlation between different multipoles, induced by the mask. The gray shaded regions indicate the bin width $\Delta \ell=25$ of our analysis.}
         \label{Fig_mixing_matrix}
   \end{figure}

The angular power spectrum of clusters in a given redshift bin $i$ can be measured from the harmonic decomposition of the observed density field. The pixelated density contrast, being defined on a 2D sphere, can be expanded in a series of spherical harmonics:
\begin{equation}        \delta_{\mathrm{cl}}^i(\mathbf{\hat{n}})=\sum_{\ell=0}^{\infty}\sum_{m=-\ell}^{m=+\ell} a_{\ell m}^i Y_{\ell m}(\mathbf{\hat{n}}), 
\end{equation}
where $Y_{\ell m}$ are the spherical harmonics, computed at the direction on the sky  $\hat{\mathbf{n}}\equiv (\theta,\varphi)$, $a_{\ell m}$ are the harmonic coefficients, defined by:
\begin{equation}
    a_{\ell m}^i= \int \mathrm{d}\mathbf{\hat{n}}\, \delta^i_{\mathrm{cl}}(\mathbf{\hat{n}}) Y^{*}_{\ell m}(\mathbf{\hat{n}}) \simeq \sum^{N_{pix}}_I \delta_{cl, I}^iY^{*}_{\ell m}(\theta_I, \varphi_I)\Delta \Omega_I,
\end{equation}
the symbol $^*$ indicates the complex conjugation operator, while $\Delta \Omega_I$ is the area of the $I$-th pixel. In this analysis, we used the angular power spectrum estimator introduced by \citet{Peebles73} and \citet{Hauser73}. 
For a partial sky survey, the masked density contrast is related to the full-sky field through a binary mask function, $\Tilde{\delta}_{\mathrm{cl}}(\mathbf{\hat{n}})=M(\mathbf{\hat{n}})\delta_{\mathrm{cl}}(\mathbf{\hat{n}})$, thus the measured pseudo-$C_\ell$ in Fig. \ref{fig_cl_measure}, named $K^{ij}_\ell$, is corrected for the sky fraction and defined as follows: 
\begin{equation}
    \label{eq_estimator_Cl}
    K^{ij}_\ell=\frac{1}{w^2_\ell}\left[\frac{1}{f_{\mathrm{sky}}(2\ell+1)}\sum_{m=-\ell}^{m=\ell} |a^i_{\ell m}a^{*j}_{\ell m}|- \frac{\Delta \Omega}{N_{\mathrm{cl}}}\delta^{ij}_K\right],
\end{equation}
where $w_\ell$ is the \textsc{Healpix} pixel window function, which removes the effect of the pixelisation, depending on the parameter $N_{\mathrm{side}}$. \\
The case $i=j$ refers to the auto power spectrum, while $i\neq j$ to the cross power spectrum between different redshift bins. We removed the shot noise contribution from the measured power spectrum, which accounts for the unclustered part of the measure, given by a discrete distribution of point-like sources. In first approximation, it depends only on the ratio $\frac{\Delta \Omega}{N_{\mathrm{cl}}}\delta^{ij}_K$, where $\delta^{ij}_K$ is the Kronecker delta, equal to zero in the cross-correlation case. Possible deviations from this term are further investigated in Sect. \ref{sect_shot_noise}. Due to the limited size of our cluster catalogues, the shot noise becomes the dominant part of the total signal at angular scales $\ell \gtrsim 100-150$, i.e. $\theta \lesssim 1.8 - 1.2$ deg, depending on the redshift bin. \\
\indent As the KiDS-DR3 catalogue does not cover the full-sky, spherical harmonics no longer provide a complete orthonormal basis to expand the angular overdensity field \citep{Camacho19}. Thus, the measured power spectrum at multipole $\ell$ depends on an underlying range of multipoles $\ell '$ \citep{Blake07}. This mode-mode coupling determines a power transfer between different multipoles and it is summarised in $R_{\ell \ell '}$, the so-called mixing matrix, which depends only on the geometry of the angular mask. The ensemble average of the measured power spectrum is related to the theoretical one through \citep{Balaguera18}: 
\begin{equation}
\label{Eq_mixing_matrix}
    \langle K^{ij}_\ell \rangle =\frac{1}{f_{\mathrm{sky}}} \sum_{\ell'} R_{\ell \ell'} C_\ell.
\end{equation}
The mixing matrix is equal to an identity matrix in full-sky surveys, where $f_{\mathrm{sky}}=1$. Starting from the survey window function, $R_{\ell \ell'}$ can be expressed in terms of the Wigner $3j$ symbols: 
\begin{equation}
    R_{\ell \ell' }=\frac{2 \ell ' +1}{4 \pi}\sum_{\ell ''} (2\ell '' +1 ) W_{\ell '' }\begin{pmatrix} \ell & \ell ' & \ell '' \\0 & 0 & 0 \end{pmatrix}  ^2, 
\end{equation}
where:
\begin{equation}
    W_\ell= \sum_{m=-\ell}^{+\ell} \frac{|I_{\ell m}|^2}{(2 \ell +1)}, 
\end{equation}
and $I_{\ell m}$ represents the spherical harmonic coefficient of the mask, given by: 
\begin{equation}
    I_{\ell m}= \int_{\Delta \Omega} Y^{*}_{\ell m}(\mathbf{\hat{n}}) \mathrm{d}\mathbf{\hat{n}} \simeq \sum^{N_{pix}}_I Y^{*}_{\ell m}(\theta_I, \varphi_I)\Delta \Omega_I. 
\end{equation}
We estimate the mixing matrix using the publicly available code \textsc{NaMaster} \citep{Alonso19}, which provides a general framework for the pseudo-$C_\ell$ analysis. The convolution function $R_{\ell^* \ell'}$ is shown in Fig. \ref{Fig_mixing_matrix}. It is peaked at $\ell^*$, while it has a drop at different multipoles, depending on the survey area and geometry. In this sense, the mixing matrix gives us an indication about the size of $\Delta \ell$ bands used to bin the measurements. Indeed, with $\Delta \ell=25$ we can keep most of the clustering signal, reducing both the effect of the window function, the size of the covariance matrix and the correlation between different bands \citep{Blake07, Balaguera18, Loureiro19}. 
For each bin we compute the weighted average:
\begin{equation}
   K^{ij}_{\Delta \ell}= \frac{\sum_{\ell \in \Delta \ell} (2\ell +1 ) K^{ij}_\ell}{\sum_{\ell \in \Delta \ell} (2\ell +1 )}. 
\end{equation} 
After the convolution with the mixing matrix, the theoretical $C_\ell$ is averaged with the same weights. As shown in Fig. \ref{fig_cl_measure}, we restrict our analysis in the range $10<\ell<175$. The lower limit reflects the validity of the Limber approximation, while the upper is a conservative value that accounts for the impact of the shot noise \citep{Balaguera18}, beyond which the clustering signal can be considered negligible.\\
The analytical error estimation adopted in \citet{Blake07}, \citet{Thomas11}, \citet{Balaguera18} and \citet{Camacho19} contains contributions from the cosmic variance and the shot noise, with a boost factor $\sqrt{f_{\mathrm{sky}}}$ \citep{ Blake07,Thomas11}. It is based on the assumption that the coefficients $a_{\ell m}$ are Gaussian distributed, while the effect of the angular window function is modelled by the parameter $f_{\mathrm{sky}}$ \citep{Blake07, Balaguera18}, resulting in a diagonal covariance matrix. These approximations do not hold in our case, due to the irregular shape of the KiDS-DR3 survey, which introduces a non-negligible correlation between different multipoles, resulting in an error leak towards other $\ell$ modes and in a reduction of the diagonal errors \citep{Crocce11}. Thus we estimate random errors directly from the dataset, using jackknife resampling. In practice, we divided our binary mask in $N_{\mathrm{JK}}=400$ non-overlapping regions, removing one of them at a time. The cluster density field and the survey area used for the angular power spectrum measurements are then updated, for every realisation, by multiplying the original \textsc{Healpix} map by the new mask.  
\subsubsection{Shot noise correction}
\label{sect_shot_noise}
\begin{figure*}[htbp]
    \centering
    \includegraphics[width=\textwidth]{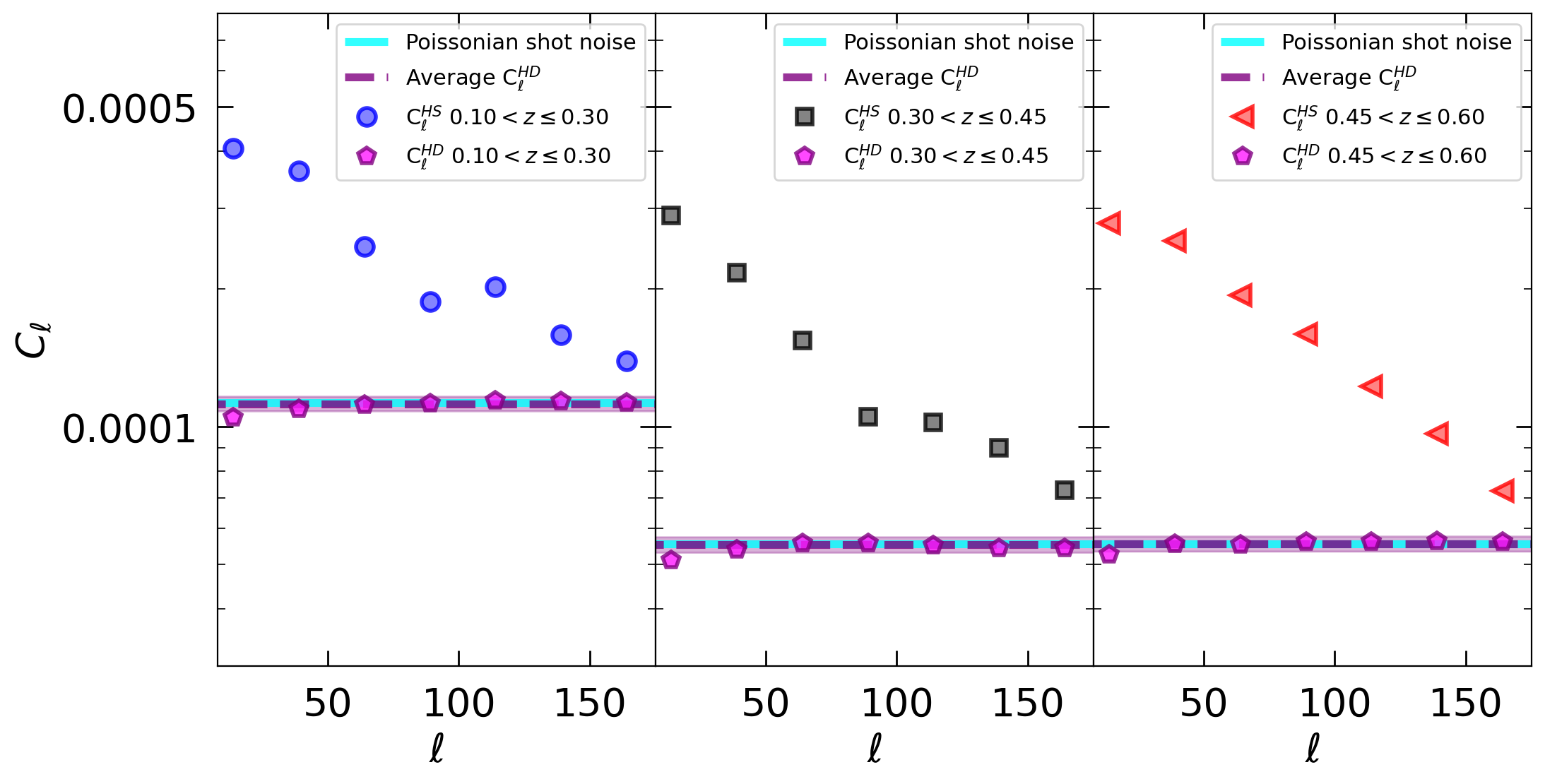}
      \caption{Shot noise estimation over 100 realisations. Blue dots, black squares and red triangles represent the angular power spectrum of the half-sum (HS) maps, which contain the contribution of both signal and noise, in our three redshift bins (panels from left to right). The purple pentagons show instead the angular power spectrum of the half-difference (HD) maps, which provides a direct estimation of the shot noise, with their average (purple dashed line) and their standard deviation (purple shaded band) in agreement with the theoretical Poissonian value (cyan solid line). }
    \label{Fig_ShotNoise}
   \end{figure*}
The quantity measured with Eq. \eqref{eq_estimator_Cl} is the sum of two contributions: the signal and the shot noise. The latter represents the unclustered part of the power spectrum caused by the discreteness of the cluster distribution. The Poisson sampling of point-like sources contributes to the auto-correlation at null separation, in real space, which brings to a constant power spectrum in harmonic space \citep{Paech17}, equal to $\frac{\Delta \Omega}{N_{\mathrm{cl}}}\delta^{ij}_K$. 
However, when dealing with real data, this simple relation may not always hold. We verify the validity of the analytical relation for the shot noise in two steps. First, we check the power spectrum of $100$ density maps derived from random cluster positions within the unmasked regions, considering the same number of cluster as in the real AMICO-KiDS catalogue. We find that the coupling with irregular mask merely increases the dispersion of the shot noise power spectrum around its theoretical prediction, with respect to full-sky surveys, without altering its mean value. Second, we expect deviations from the Poissonian shot noise due to halo exclusion and nonlinear
effects \citep{Giocoli10, Baldauf13}. The exclusion simply consists in the fact that clusters cannot overlap, i.e. the distance between clusters cannot be smaller than the sum of their radii, $R$ \citep{Baldauf13, Paech17}. In other words, the cluster-cluster real-space correlation function $\xi_{\mathrm{cc}}(r)=-1$, for $r<R$, which means that the probability to find another cluster is zero \citep{Baldauf13}. The exclusion introduces a mass-dependent deviation from the Poissonian shot noise, since objects with different mass occupy different fractions of the sampling volume, while the nonlinear clustering effectively increases the noise term \citep{Paech17}.\\
\indent To test the robustness of the Poissonian shot noise hypothesis at the angular scales of our interest, we need a pratical way to disentangle signal and noise. Following \citet{Ando18}, \citet{Makiya18} and \citet{Ibitoye22}, we first randomly divide the catalogue into two submaps, $\delta_{1,\mathrm{cl}}$ and $\delta_{2,\mathrm{cl}}$, both of which contain roughly the same number of clusters. Then we build the half-sum, $\mathrm{HS}=\frac{1}{2}(\delta_{1,\mathrm{cl}}+\delta_{2,\mathrm{cl}})$, and the half-difference density fluctuation maps, $\mathrm{HD}=\frac{1}{2}(\delta_{1,\mathrm{cl}}-\delta_{2,\mathrm{cl}})$. By construction, the former contains both signal and noise, while in the latter the signal cancels out, leaving only the shot noise contribution. Since the division of the catalogue into subsets is a random process, the estimated HD map and its power spectrum $C_\ell^\mathrm{HD}$ slightly change for different realisations \citep{Ando18}. Thus we averaged over 100 realisations, finding that the shot noise approximation holds in every redshift bin for all the multipoles considered in our analysis. In Fig. \ref{Fig_ShotNoise} we show the ensamble average of $C_\ell^{\mathrm{HS}}$ and $C_\ell^\mathrm{HD}$ power spectra, and the comparison with the Poissonian shot noise. Due to the very low number of clusters in our redshift bins, following \citet{Loureiro19} we include an extra shot noise term, such that $C^{\mathrm{th}, i}_{\Delta\ell}\rightarrow C^{\mathrm{th}, i}_{\Delta\ell} +\mathcal{S}^i$. The $\mathcal{S}^i$ nuisance parameters are forward modelled at the likelihood level, allowing them to vary within a Gaussian prior given by a mean equal to zero and the same standard deviation of $C_\ell^\mathrm{HD}$, as shown in Table \ref{table_par}.
\subsection{The angular power spectrum model}
\begin{figure*}[htbp]
    \centering
    \includegraphics[width=\textwidth]{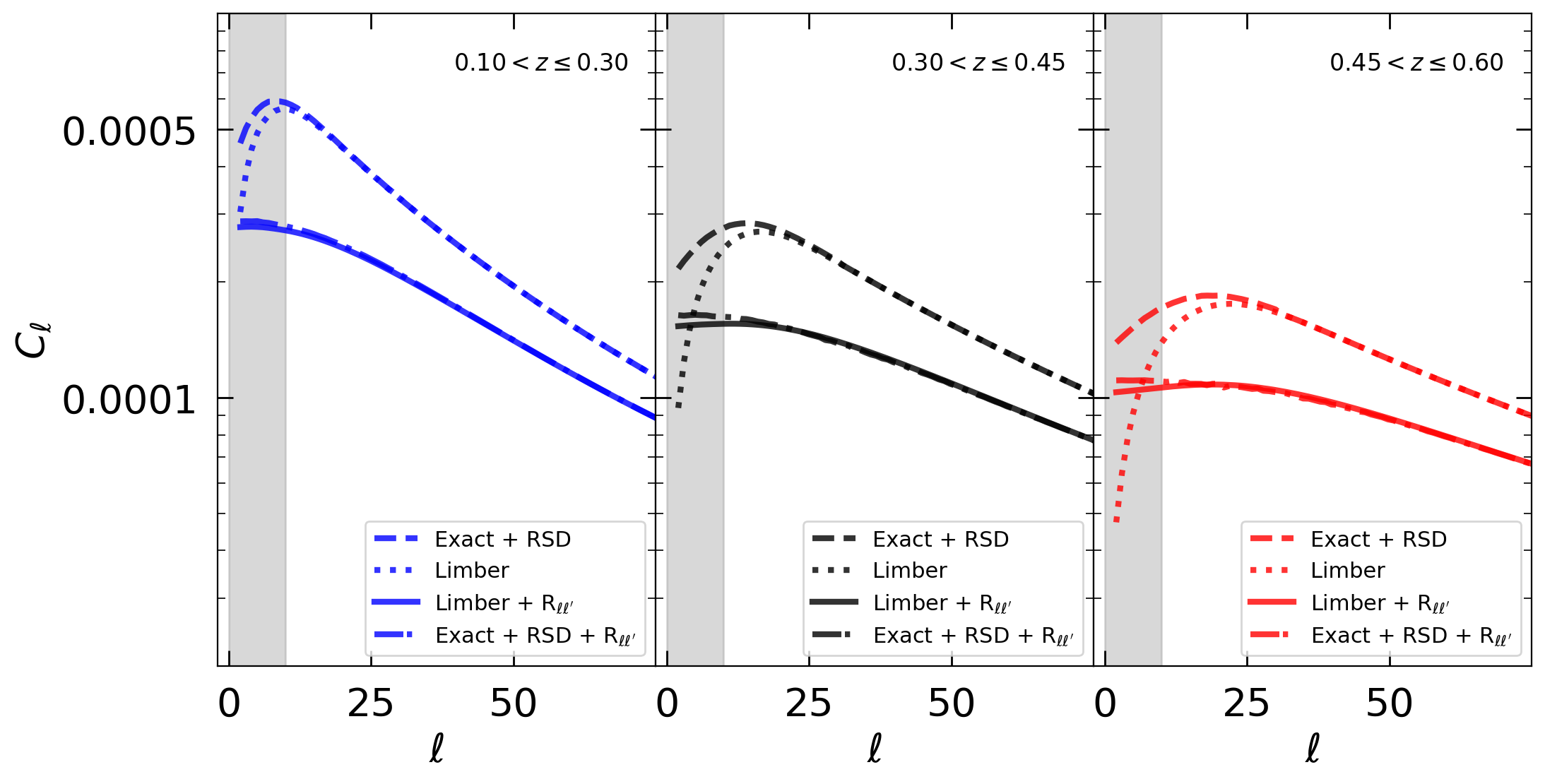}
      \caption{Theoretical angular power spectrum for our three redshift bins. The Limber approximation is represented with a dotted line, while the exact computation presented in Eqs. \eqref{eq_window_real_space} and \eqref{eq_window_redshift_space} with a dashed line. The convolution with the mixing matrix is shown, respectively, with a solid line (Limber approximation, i.e. the model used in our analysis), and with a dot-dashed line (exact computation). After the mode-mode coupling, the Limber approximation deviates only for $\ell \lesssim 10$, i.e. for the angular range already excluded from our analysis (gray shaded region).}
         \label{Fig_Limber_approximation}
   \end{figure*}
The angular power spectrum is modelled from the spatial power spectrum through a projection kernel, which takes into account redshift evolution and radial selection effects. Its exact computation is given by \citep{Padmanabhan07, Thomas11, Asorey12, Camacho19}: 
\begin{equation}
    C_{\ell}^i = \frac{2}{\pi} \int \, \mathrm{d}k \, k^2 P_{\mathrm{DM}}(k) \left[ \Psi_{\ell}^i(k) +\Psi^{i, r}_\ell(k) \right]^2.
\end{equation}
The kernel function $\Psi_{\ell}^i(k)$ describes the mapping of $k$ to $\ell$ in real space and is defined as:
\begin{equation}
\label{eq_window_real_space}
    \Psi_{\ell}^i(k)=\int \, \mathrm{d}z \, b(z) \phi^i(z) D(z) j_\ell(kr(z)),
\end{equation}
where $\phi^i(z)$ and $b(z)$ are computed with Eqs. \eqref{eq_selection_function} and \eqref{eq_bias_eff}, respectively, $j_\ell(x)$ is the spherical Bessel function and $D(z)$ is the linear growth factor, normalised such that $D(0)=1$ \citep{Camacho19}. The second term incorporates the linear Kaiser effect \citep{Padmanabhan07, Asorey12}, i.e. the enhancement in 3D power spectrum due to cluster peculiar velocities:
\begin{equation}
\begin{split}
\label{eq_window_redshift_space}
    \Psi^{i,r}(k)= \int \mathrm{d}z \phi^i(z) f(z) D(z) \biggl[ \frac{2\ell^2+2\ell-1}{(2\ell +3)(2\ell -1)} j_\ell(kr(z))\\ - \frac{\ell (\ell-1)}{(2\ell -1)(2\ell +1 )}j_{\ell-2}(kr(z)) -  \frac{(\ell+1)(\ell+2)}{(2\ell+1)(2\ell+3)}j_{\ell+2}(kr(z)) \biggr]. 
\end{split}
\end{equation}
Cluster peculiar velocities are smaller with respect to galaxy ones, thus they have a minor effect in our broad redshift selection function, and RSD are erased by the radial projection \citep{Padmanabhan07}. In particular, for $\ell \gg 0$, the term $\Psi^{i,r}$ tends to zero, so that the total window function reduces to Eq. \eqref{eq_window_real_space} \citep{Padmanabhan07,Thomas11}. However, since the evaluation of spherical Bessel functions is still quite computationally demanding, we make use of the Limber approximation \citep{Limber53}:\begin{equation}
    C_\ell^{ij}= b^i_{\mathrm{eff}} b^j_{\mathrm{eff}} \int_0^\infty \mathrm{d}z \phi^i(z)\phi^j(z) P_{\mathrm{DM}}\left(\frac{\ell+\frac{1}{2}}{r(z)}, z\right) \frac{H(z)}{r^2(z)c},
\end{equation}
where $H(z)$ is the Hubble parameter and the photo-$z$ effects are included through the radial selection function, $\phi(z)$ \citep{Asorey12}, see Sect. \ref{Selection function}. Here we underline that $P_{\mathrm{DM}}(k,z)=P_{\mathrm{DM}}(k)D(z)^2$ is strictly valid only in linear theory \citep{Blake07}.\\
\indent Finally, there are two ways to consider the mode-mode coupling induced by the mask. One can solve the linear system in Eq. \eqref{Eq_mixing_matrix}. This requires to bin the pseudo-$C_\ell$ into even more larger bandpowers, since it is computationally expensive and unstable to deconvolve the effect of the mixing matrix from our noisy data \citep{Oliveira21}. For this reason, we decide to include the angular selection directly at the level of the likelihood analysis, choosing the forward modelling \citep{Balaguera18, Loureiro19, Xavier19}. \\
\indent In Fig. \ref{Fig_Limber_approximation} we show how redshift-space distortions and partial sky convolution can alter the shape of the angular power spectrum. In particular, the effect of the Limber approximation can be noted only for $\ell \lesssim 10$, i.e. for multipoles already excluded from our analysis. On the other hand, modifications due to the mixing matrix affect much smaller scales ($\ell \lesssim 150$), thus we need to properly include them in our model.
\section{Cosmological analysis}
\label{Results}

\begin{figure*}[htbp]
\centering
\includegraphics[width=\textwidth]{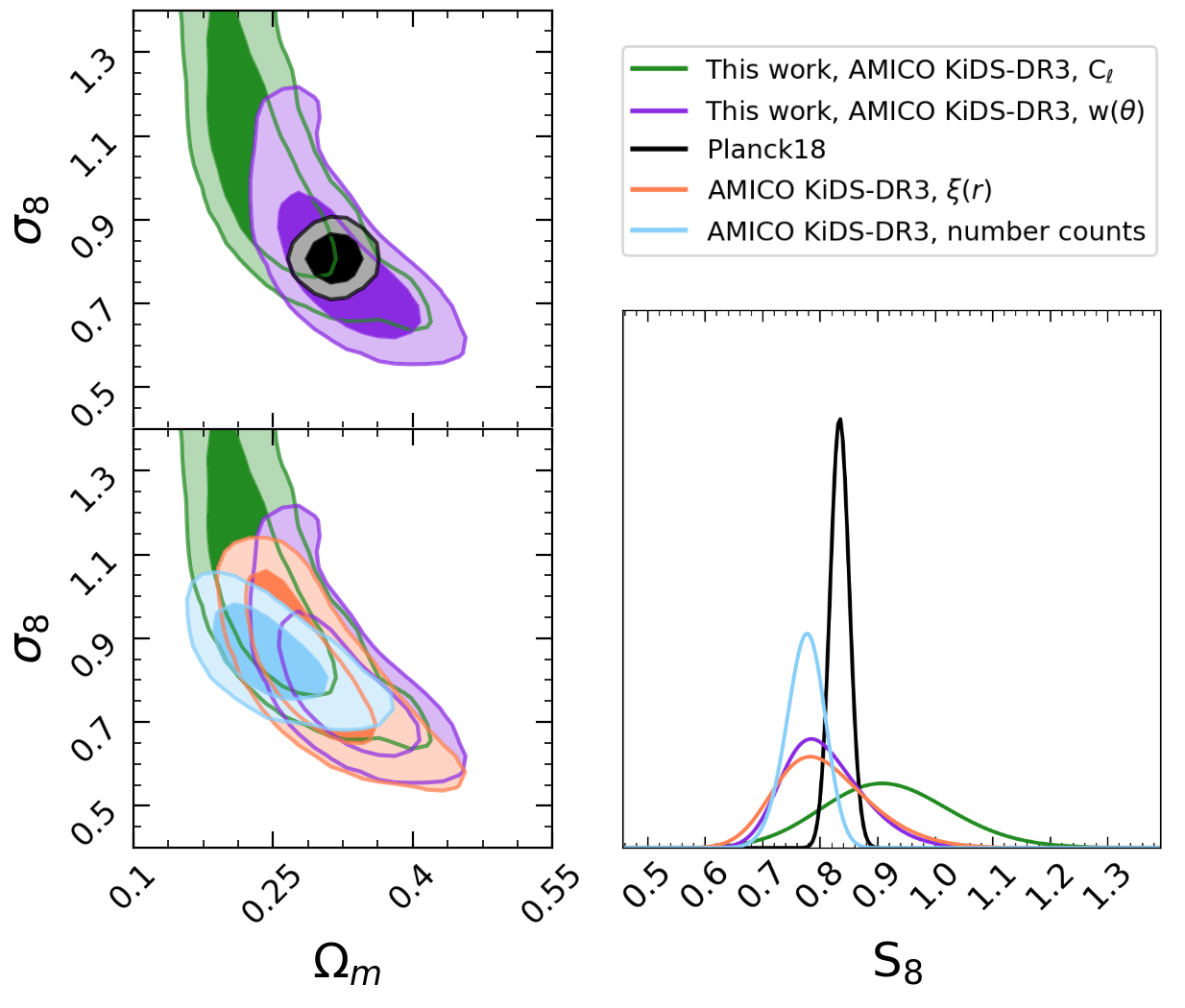}
\caption{Cosmological constraints in the $\Omega_{\mathrm{m}}-\sigma_8$ plane, with $68\%$ and $95\%$ confidence intervals, obtained considering angular power spectrum (green contours) and correlation function (purple contours). Our findings are compared to the cosmological constraints derived from \citet{Planck20} (top left) and from the same KiDS-DR3 cluster catalogue (bottom left), in particular to the number counts and 3D correlation function analyses presented in \citet{Lesci_counts, Lesci22}. Right: summary plot containing the 1D marginalised posteriors for the structure growth parameter, $S_8$.}
\label{Corner}
\end{figure*}

\subsection{Likelihood}
The analyses of the angular correlation function and power spectrum are performed through Bayesian statistics. We estimate the set of cosmological parameters in Table \ref{table_par}, which enter the model $m$, by adopting a Gaussian likelihood: 
\begin{equation}
\mathcal{L}_k\propto \mathrm{exp}(-\chi^2_k/2),
\end{equation}
in the $k$-th redshift bin, where: 
\begin{equation}
    \chi^2_k=\sum_{i=1}^N \sum_{j=1}^N (\mu^{d}_i-\mu^m_i)_{(k)} C^{-1}_{i,j,(k)} (\mu^{d}_j-\mu^m_j)_{(k)},
\end{equation}
where $N$ is the number of angular bins, $\mu$ is the correlation statistic involved, i.e. angular correlation function or power spectrum, and the superscript $d$ and $m$ refer to the quantities obtained from the data and computed with the model, respectively. $C^{-1}_{i,j,(k)}$ is the inverse of the covariance matrix in the $k$-th redshift bin, estimated directly from the data for $N_{\mathrm{JK}}$ jackknife resamplings \citep[for example][]{Norberg09}:
\begin{equation}
    C_{i,j}=\frac{N_{\mathrm{JK}}-1}{N_{\mathrm{JK}}}\sum_{\ell=1}^{N_{\mathrm{JK}}}(\mu^\ell_i-\overline{\mu}_i)(\mu^\ell_j-\overline{\mu}_j),
\end{equation}
with expectation values $\overline{\mu}_i=\sum_{\ell=1}^{N_\mathrm{res}} \mu_i^{\ell}/N_{\mathrm{JK}}$. We neglect correlations between different redshift slices, so that the total likelihood is simply given by the product of the individual likelihoods, $\mathcal{L}_k$.
\subsection{Cosmological results}   
\begin{figure}[]
\centering
\includegraphics[width=\hsize]{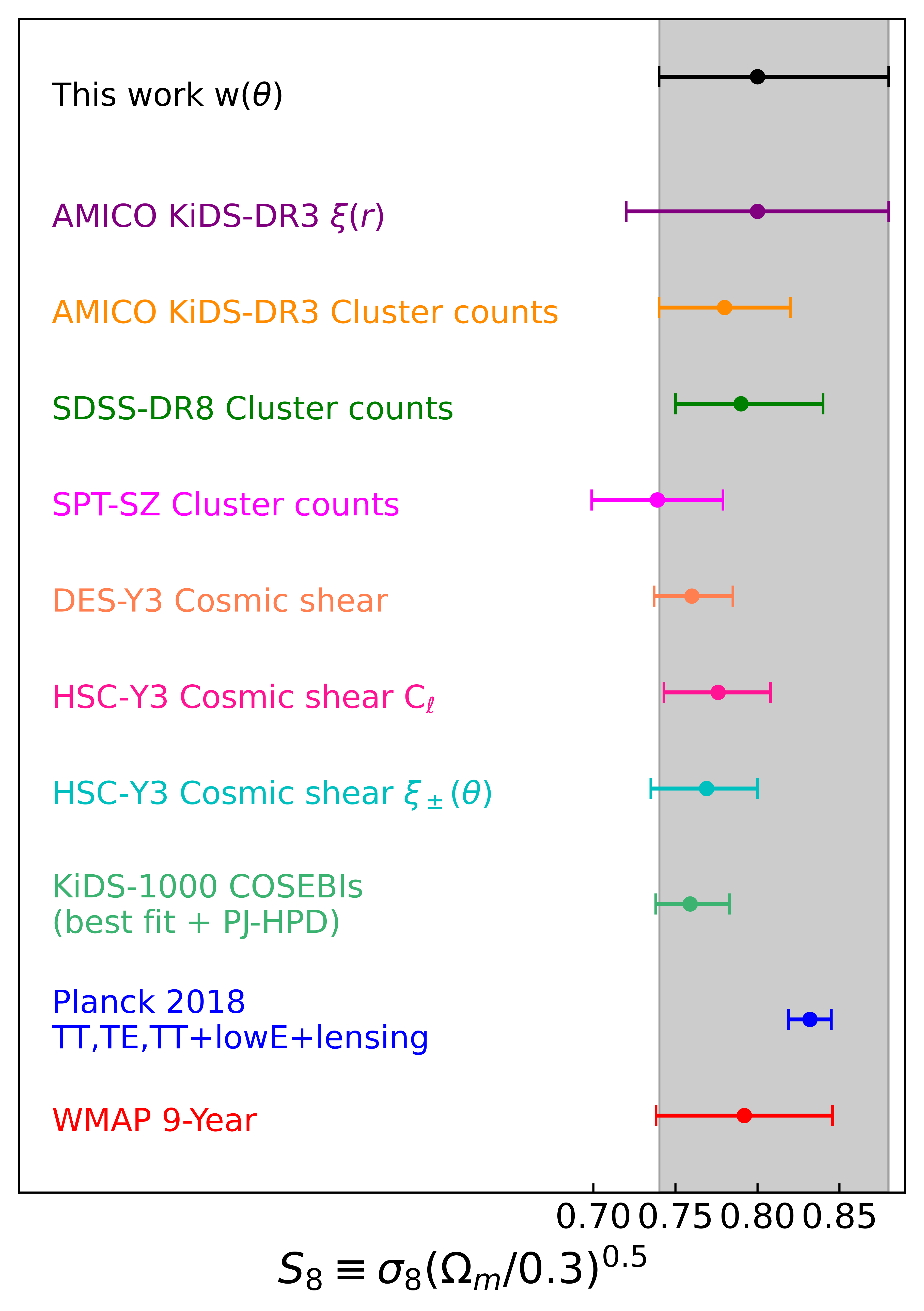}
  \caption{Comparison of the constraints on $S_8$, given by the posterior median, the 16-th and the 84-th percentiles. From top to bottom: angular cluster clustering in the AMICO KiDS-DR3 catalogue obtained in this work (black dot); 3D cluster clustering from \citet[][purple dot]{Lesci22}; cluster counts from \citet[][orange dot]{Lesci_counts}, \citet[][green dot]{Costanzi19} and \citet[][magenta dot]{Bocquet19}; cosmic shear from \citet[][coral dot]{Amon22, Secco22}, \citet[][pink dot]{Li023}, \citet[][cyan dot]{Dalal23} and \citet[][pale green dot]{Asgari21}; CMB results from \citet[][blue dot]{Planck20} and \citet[][red dot]{WMAP13}.}
  \label{Fig_comparison_S8}

\end{figure}

The Bayesian analysis is performed by adopting uniform priors on both $\Omega_{\mathrm{m}}$ and $\sigma_8$, while we assume Gaussian priors around the mean values from \citet[][Table 2, TT, TE and EE+lowE]{Planck20} for the baryon density, $\Omega_\mathrm{b}$, the primordial spectral index, $n_\mathrm{s}$ and the normalised Hubble constant, $h$. They are summarised in Table \ref{table_par}. We use Gaussian priors also for $\alpha$, $\beta$, $\gamma$, $\sigma_{\mathrm{intr},0}$ and $\sigma_{\mathrm{intr},\lambda^*}$, the parameters of the mass-richness scaling relation, whose medians and standard deviations are taken from the posterior distribution of the cluster counts and weak lensing joint analysis, as derived by \citet{Lesci_counts}.\\
\indent In Fig. \ref{Corner} we show the results of the Monte Carlo Markov Chain (MCMC) cosmological analysis, in the $\Omega_{\mathrm{m}}-\sigma_8$ plane. For the angular correlation function we find $\Omega_{\mathrm{m}}=0.32^{+0.05}_{-0.04}$ and $\sigma_8=0.77^{+0.13}_{-0.09}$ as the medians, the $16$th and the $84$th percentiles of the marginalised 1D posterior distributions. The constraint on the structure growth parameter, $S_8=0.80^{+0.08}_{-0.06}$, is presented in Fig. \ref{Fig_comparison_S8} as our main outcome, and compared with several previous studies available in the literature. In particular, we find a $1\sigma$ agreement with cosmological constraints from Wilkinson Microwave Anisotropy Probe \citep[WMAP,][Table 3, WMAP-only, Year 9]{WMAP13} and \textit{Planck} \citep[][Table 2, TT, TE, and EE+lowE]{Planck20}. An equivalent level of agreement is found also with the number counts analysis presented in \citet{Lesci_counts} using the same AMICO KiDS-DR3 cluster sample, in \citet{Costanzi19} based on SDSS-DR8 cluster data, in \citet{Bocquet19} with the 2500 deg$^2$ South Pole Telescope - Sunyaev-Zel'dovich (SPT-SZ) survey data and with constraints from cosmic shear in DES Year 3 \citep{Amon22, Secco22}, HSC Year 3 \citep{Dalal23, Li023} and KiDS-DR4 \citep{Asgari21}. Instead, we note a $1\sigma$ tension with $S_8$ from cluster abundances and weak lensing in DES Year 1 data \citep{Abbott20}, probably related to richness-dependent effects, since it significantly reduces when their sample is limited to clusters with $\lambda^* \geq 30$. \\
\indent Concerning cluster clustering, our main achievement shows the competitiveness of the 2D correlation function with respect to its 3D counterpart. Indeed, the current study provides more constraining power than the 3D correlation case discussed in \citet{Lesci22}, where the measurement of $\xi(r)$ is performed within two redshift bins, $0.1\leq z\leq0.3$ and $0.35\leq z\leq 0.6$. This is partially due to the slightly larger sample considered, with 228 more clusters, and highlights the importance of the tomographic strategy adopted in photometric redshift surveys, for which several parameters, such as the bin width, the photometric redshift error and the number density of detections, need to be balanced, as discussed in Sect. \ref{Data}. Our tighter constraints are also confirmed by repeating the full MCMC analysis over 1000 bootstrap resampling, with replacement, which yields to $\Omega_{\mathrm{m}}=0.33^{+0.04}_{-0.04}$, $\sigma_8=0.75^{+0.11}_{-0.08}$ and $S_8=0.79^{+0.08}_{-0.05}$, in excellent agreement with our findings.\\
\indent The angular power spectrum analysis includes an extra shot noise parameter for each redshift bin, with a Gaussian prior derived with the methodology described in Sect. \ref{sect_shot_noise}. As shown in Fig. \ref{Corner} the angular correlation function and the angular power spectrum produce statistically consistent results, although the latter exhibits a much lower constraining power, with wider constraints in particular on $\sigma_8$. This is probably due to the relative importance of the shot noise, which equals the signal contribution even at sub-degree scales and prevents us from extending our angular range to $\ell \gtrsim 175$. We found $\Omega_{\mathrm{m}}=0.24^{+0.05}_{-0.04}$, in excellent agreement with $\Omega_{\mathrm{m}}=0.24^{+0.03}_{-0.04}$ presented in \citet{Lesci_counts} with number counts, using the same cluster catalogue, and $\sigma_8=1.01^{+0.25}_{-0.17}$, which yield to $S_8=0.93^{+0.11}_{-0.12}$, consistent within 1$\sigma$ with both WMAP and \textit{Planck} constraints.
As shown in \citet{Sartoris16} and \citet{Garrel22}, the combination of cluster clustering with the more constraining cluster number counts is particularly important since it can highly improve the parameter estimation accuracy both in the $\Lambda$CDM $\Omega_m-\sigma_8$ plane and in the dynamical dark energy $w_0-w_\mathrm{a}$ plane. The full list of parameters, with prior intervals, the medians, the $16$th and the $84$th percentiles of the marginalised posterior distributions are shown in Table \ref{table_par}. \\
\indent Finally, we perform some tests to confirm the robustness of our results. First, we verify that our findings are stable if we adopt the halo mass function parameters provided by \citet{Despali16} in the model of the cluster redshift distribution. Second, we focus on the inspection of Figs. \ref{fig_wtheta_measure} and \ref{fig_cl_measure}, which reveals an excess of clustering in the redshift range $0.45<z\leq0.60$, common both to $w(\theta)$ and $C_\ell$, with respect to the model median prediction. This tension is marginal at $2\sigma$ and has a negligible impact on the overall conclusion. We check our constraints by repeating the entire MCMC analysis, in the first and second redshift bins only. We found consistent results, but with a larger uncertainty due to the lower statistics, since we excluded approximately the $40\%$ of the clusters. In particular, the exclusion of the third bin causes a general broadening of the posterior distributions, with a slight shift for $\Omega_{\mathrm{m}}$, to values of $0.25^{+0.07}_{-0.04}$ and $0.34^{+0.07}_{-0.05}$, for angular power spectrum and correlation function, respectively. These results are fully consistent with our expectations. We underline that the excess of clustering appears to be independent of the selection effects, since it remains even if we select more massive clusters with $\lambda^*\geq20$, for which both the purity and the completeness are higher \citep{Maturi19}. The underestimation in the model might stem from a mismatch between the recovered and the true cluster masses derived from weak lensing calibration, which propagates in the effective bias of the cluster sample. Alternatively, inaccuracies in the cluster selection function, especially at higher redshifts where the cluster redshift distribution deviates more from the theoretical one, could also contribute. However, these systematics are not fundamentally inconsistent with our findings, as we show repeating the analysis without the last redshift bin, and with previous cosmological results obtained in \citet{Lesci_counts, Lesci22}, using number counts and 3D clustering. We aim at better investigating this aspect by exploiting the larger statistics offered by the KiDS-DR4 \citep{Kuijken19} data.
\section{Conclusions}
\label{Conclusions}
In this paper we presented the cosmological constraints derived from the angular clustering properties of the KiDS-DR3 cluster catalogue. The sample of clusters, which has been constructed with the AMICO algorithm, consists of 5162 galaxy clusters with intrinsic richness $\lambda^*\geq15$. Using a tomographic approach, we measured the angular correlation function and power spectrum in three different photometric redshift bins, $z \in (0.10, 0.30]$, $z \in (0.30, 0.45]$ and $z \in (0.45, 0.60]$, whose widths were selected in order to balance the statistics and the photometric errors. For the angular power spectrum, we verified that the Poissonian shot noise approximation holds in every redshift bin and for all the multipoles considered in our study. \\
\indent We modelled the clustering signal by taking into account the effects of the photometric errors on the redshift selection function, and considering the mass-richness scaling relation from the weak lensing analysis by \citet{Lesci_counts}, in order to estimate the effective bias and the redshift distribution of the cluster sample. For the first time, we found cosmological constraints from the angular correlation function and power spectrum of a photometric-redshift cluster catalogue. From the MCMC analyses, we obtained $\Omega_{\mathrm{m}}=0.32^{+0.05}_{-0.04}$, $\sigma_8=0.77^{+0.13}_{-0.09}$ and $S_8=0.80^{+0.08}_{-0.06}$ for $w(\theta)$, and $\Omega_{\mathrm{m}}=0.24^{+0.05}_{-0.04}$, $\sigma_8=1.01^{+0.25}_{-0.17}$ and $S_8=0.93^{+0.11}_{-0.12}$ for $C_\ell$. Both exhibit a $1\sigma$ agreement with the literature results reported in Fig. \ref{Corner}, which includes different cosmological probes like CMB, cluster number counts and cluster clustering. From a comparison with \citet{Lesci22}, our work has shown that the 2D clustering from a photometric-redshift survey can provide competitive constraints with respect to the full 3D clustering, with the advantage that our findings are cosmological independent, since they rely on the cluster angular positions alone, without any cosmological assumption in converting redshifts to distances. Indeed, from the angular correlation function we derived tighter uncertainties based on the same AMICO KiDS-DR3 cluster catalogue, but with a slightly larger sample. This fact reveals the importance of the tomographic strategy adopted in order to fully exploit the cosmological information contained in the cluster catalogue. On the other hand, the angular power spectrum yielded to a wider posterior, in particular with regard to the parameter $\sigma_8$, and does not allow to cover the full angular range explored with $w(\theta)$, due to the relative importance of the shot noise. \\
\indent We tested the the robustness of our study with respect to the parameterisation of the halo mass function presented in \citet{Despali16}. Moreover, we detected an excess of clustering in the redshift range $0.45<z\leq0.60$, which does not depend on the selection in richness, since it remains even if we consider only clusters with $\lambda^* \geq 20$. However, this tension is marginal at $2\sigma$ and does not affect our results, which are stable even if we exclude the third redshift bin, restricting our redshift range to $(0.10,0.45]$. \\
\indent We expect more stringent constraints from the analyses of KiDS-DR4 \citep{Kuijken19}, which covers an area of approximately 1000 deg$^2$ and includes the photometry of the VISTA Kilo-degree INfrared Galaxy survey \citep[VIKING; see][]{Edge13}, and of the final KiDS-DR5 \citep{Wright23}, which will contain data from the full $1350$ deg$^2$ of the KiDS/VIKING footprint. They will allow us to better investigate the differences, the behaviour and the benefits of these two complementary statistics, as well as their combination with number counts and other independent cosmological probes. In the future, we expect an extensive use of the angular clustering of galaxy clusters within the next-generation photometric redshift surveys, like for example \textit{Euclid} \citep{Laureijs11, Scaramella14, Amendola18, Scaramella22}, which will allow us to constrain the parameters of the Dark Energy equation of state, leading to significant advances in the field of the observational cosmology. 

%
\begin{acknowledgements}
     We would like to thank K. Paech, N. Hamaus, J. Weller and S. Hagstotz for the constructive conversations about angular clustering. 
     We thank Joachim Harnois-Déraps for the valuable comments that enriched the publication. We acknowledge support from the grants PRIN-MIUR 2017 WSCC32 and ASI n.2018-23-HH.0, and the use of computational resources from the parallel computing cluster of the Open Physics Hub (\url{https://site.unibo.it/openphysicshub/en}) at the Physics and Astronomy Department in Bologna. GC acknowledges the support from the grant ASI n.2018-23-HH.0. MR acknowledges the support from the INAF mini-grant 2022 "GALCLOCK". 
\end{acknowledgements}

%
%

\citestyle{aa}
\bibliographystyle{aa}
\bibliography{bibliografia}
%




   
  



\end{document}